\documentclass[12pt]{article}

\usepackage{latexsym}
\usepackage{amssymb}
\usepackage{amsfonts}
\usepackage{graphicx}
\usepackage[usenames]{color}

\pdfoutput=1

\def\tr{\,{\rm Tr\,}}
\def\bj{{\bar\jmath}}
\def\CA{{\cal A}}\def\CH{{\cal H}}\def\CT{{\cal T}}
\def\CC{{\cal C}}\def\CD{{\cal D}}

\def\CO{{\cal O}}\def\CN{{\cal N}}
\def\CV{{\cal V}}\def\CI{{\cal I}}
\def\be{\begin{equation}}
\def\ee{\end{equation}}
\def\IC{\mathbb{C}}\def\IZ{\mathbb{Z}}
\def\slh{\widehat{sl}}

\def\ov{\overline}

\begin{document}

\begin{titlepage}
\begin{center}

\vskip 1.5 cm

{\LARGE\bf A-D-E Classification}
\vskip .3 cm
{\LARGE\bf of Conformal Field Theories}

\vskip .5 cm

Andrea CAPPELLI \\
{\em INFN \\ Via G. Sansone 1, 50019 Sesto Fiorentino (Firenze), Italy}
\\
\vskip .3 cm

Jean-Bernard ZUBER\\
{\em LPTHE, Tour 24-25, 5{\`e}me {\'e}tage,
\\ Universit\'e Pierre et Marie Curie –- Paris 6,
\\ 4 Place Jussieu, F 75252 Paris Cedex 5, France
}

\end{center}
\vskip .5 cm

\begin{abstract}

The ADE classification scheme is encountered in many areas of
mathematics, most notably in the study of Lie algebras. 
Here such a scheme is shown to describe families of 
two-dimensional conformal field theories.

\end{abstract}

\vskip .5 cm

\begin{center}
{\it Review article to appear in Scholapedia, http://www.scholarpedia.org/}
\end{center}

\vfill
\end{titlepage}

\pagenumbering{arabic}

\section{ Overview}
 
Conformal Field Theories (CFT) in two space-time dimensions have been
the object of intensive work since the mid eighties, after the
fundamental paper \cite{BPZ84}.
These theories have very diverse and important physical applications,
from the description of critical behavior in statistical mechanics and
solid state physics in low dimension, to the worldsheet description of
string theory.
Furthermore, their remarkable analytical and algebraic structures and
their connections with many other domains of mathematics make them
outstanding laboratories of new techniques and ideas.
A particularly striking feature is the possibility of classifying
large classes of CFT through the study of representations of the
Virasoro algebra of conformal transformations \cite{CFT}.

The classification of CFT partition functions, leading to the ADE 
scheme, follows from exploiting the additional symmetry
under modular transformations, the discrete coordinate changes
that leave invariant the double-periodic finite-size geometry of the torus.

\begin{figure}[t]
\begin{center}
\includegraphics[width=5cm]{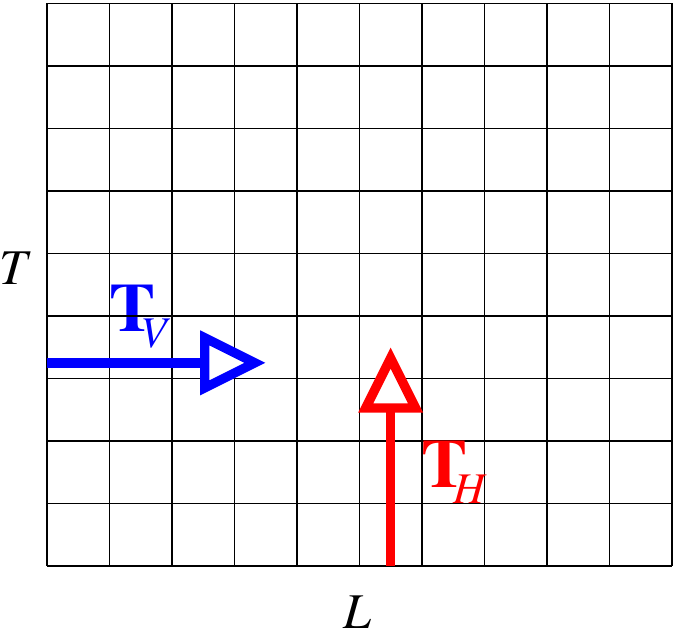}
\end{center}
\caption{Two transfer matrices of a statistical lattice model.}
\label{fig-tm}
\end{figure}

Modular invariance has a precursor in ordinary statistical mechanics 
of lattice models, like the Ising model \cite{Ising}.   
Consider a finite $L\times T$ square lattice in two dimensions as in
Fig.\ref{fig-tm}: it is common to use the transfer matrix formalism to
compute the partition function and other quantities of physical
relevance in the model.
Each configuration of degrees of freedom  on a row 
of the lattice is associated with a state vector in a Hilbert space. 
The transfer matrix ${\bf T}_H$ is an operator acting on  a state vector 
and manufacturing the state vector of the next row. 
It may thus be regarded as a discrete version of a time evolution operator, 
for time flowing vertically, see Fig. 1.  
The partition function of a system with $T$ rows and periodic 
boundary conditions in the time direction is then 
$Z= {\rm tr} {\bf T}_H^T$.
Now the model also admits a transfer matrix ${\bf T}_V$ in the
orthogonal direction, (a column-to-column transfer matrix) and if
periodic boundary conditions are also imposed in that direction, one
has an alternative expression of the partition function: 
$ Z={\rm \tr} {\bf T}_V^L$.  
Therefore, in the doubly-periodic toroidal geometry
there are two ways of writing the partition function.  Imposing their
equality may give some useful informations on the model, but in
general these are too weak to constraint its operator content,
spectrum, etc.

In the case of conformal field theories, there are special features
that allow for a complete solution of the modular invariance conditions
\cite{CFT}:
\begin{itemize}
\item 
The spectrum of states in the theory, is organized in families,
  the ``conformal towers'', with an infinite dimensional algebra
  acting as a spectrum generating algebra, (the Virasoro algebra or
  one of its extensions \cite{Vir}) \cite{BPZ84}.
The structure of these conformal towers is given by the 
representation theory of the ``chiral'' algebra parameterized by 
the Virasoro central charge $c$  and other quantum numbers.
\item 
  the Hamiltonian can be expressed in terms of the Virasoro generator of
  dilatations $L_0$ and its conjugate $\bar L_0$, whose eigenvalues give the
  conformal weights (scale dimensions).  The lowest-energy state in each
  tower, i.e. in each Virasoro representation, is called the ``highest weight
  vector'' with conformal weight $h$ and the excited states are obtained by
  acting on it with algebra generators.  (Note the mismatch of terminology:
  the highest weight vector is actually the lowest energy state.)
The structure of each tower is encoded in the character of the
representation, $\chi_h(q)$, which is the generating function of the
dimensions of eigenspaces of given energy (see below eq. (\ref{chara})).
\item 
The partition function $Z$ can be rewritten in terms of the Hamiltonian 
and thus of the Virasoro generators $L_0,\bar L_0$, 
(eq. (\ref{evol}) hereafter). 
As a consequence, it involves a sum of bilinears of characters with 
undetermined non negative integer multiplicities of representations 
in the theory, see eq. (\ref{toruspf}). 
\item 
 The requirement of modular invariance constrains these multiplicities
and allows for the complete classification of partition functions 
in some theories for which that character expansion of $Z$ is finite.
\end{itemize}

In this contribution, we describe this classification program 
in the simplest classes of conformal theories, namely the Virasoro 
minimal models, having central charge $c<1$, and the models with 
the affine Lie algebra $\slh (2)$ as an extended symmetry.
The main result is that the modular invariant partition 
functions are in one-to-one correspondence with the Dynkin diagrams of 
the A, D and E types \cite{Hum72}.
Each partition function defines an independent theory with specific
Hilbert space and field content.


\section{Modular invariant partition functions}

Two-dimensional conformal field theories are quantum field theories
enjoying covariance properties under conformal, {\it i.e.} local scale, 
transformations \cite{CFT}.
It is postulated that such theories exist and are consistent 
on any two-dimensional Riemann surface \cite{Ver}. 
In the case of the plane with complex coordinate $z$, one shows that 
$z$ and its complex conjugate $\bar z$ may be treated as independent 
variables, called holomorphic and anti-holomorphic coordinates. 
Infinitesimal conformal transformations are generated by two copies 
of the Virasoro algebra \cite{Vir}, acting on $z$ and $\bar z$
and called right and left chiral algebras, respectively.  
More generally, one may consider theories with an extended chiral
algebra $\CA$ containing Virasoro, such as the affine Lie
({Kac--Moody}) algebras \cite{KM}, the superconformal algebras, etc.
The Hilbert space $\CH$ of the CFT decomposes onto pairs of representations
$\CV_j$ of the left and right copies of the Virasoro algebra or of
$\CA$.  For ``rational'' conformal field theories (RCFT), the number
of such irreducible representations $\CV_j$ is finite.
We label by $j\in {\cal I}$ and $\bar\jmath\in {\cal I}$
the left and right irreducible representations, respectively.
These pairs of representations are in one-to-one correspondence with 
the ``primary conformal fields'' $\phi_{j\bar \jmath}$ \cite{CFT}.

The finite decomposition of the RCFT Hilbert space can be written:
\begin{equation}
\CH =\oplus_{j \bar\jmath\in {\cal I}} \
N_{j\bar\jmath}\ {\cal V}_j\otimes {\cal V}_{\bar\jmath}\ , 
\label{hilbert}
\end{equation}
where $N_{j\bar\jmath}$ are non negative integer multiplicities.
These are subjected to consistency constraints, 
due to the fact that the RCFT must exist and be consistent
on any Riemann surface \cite{Ver}. 
In particular, a crucial condition on the torus is the modular invariance
of the partition function: this requirement determines 
the $N_{j\bar\jmath}$ as described hereafter.

\begin{figure}[t]
\begin{center}
\includegraphics[width=11cm]{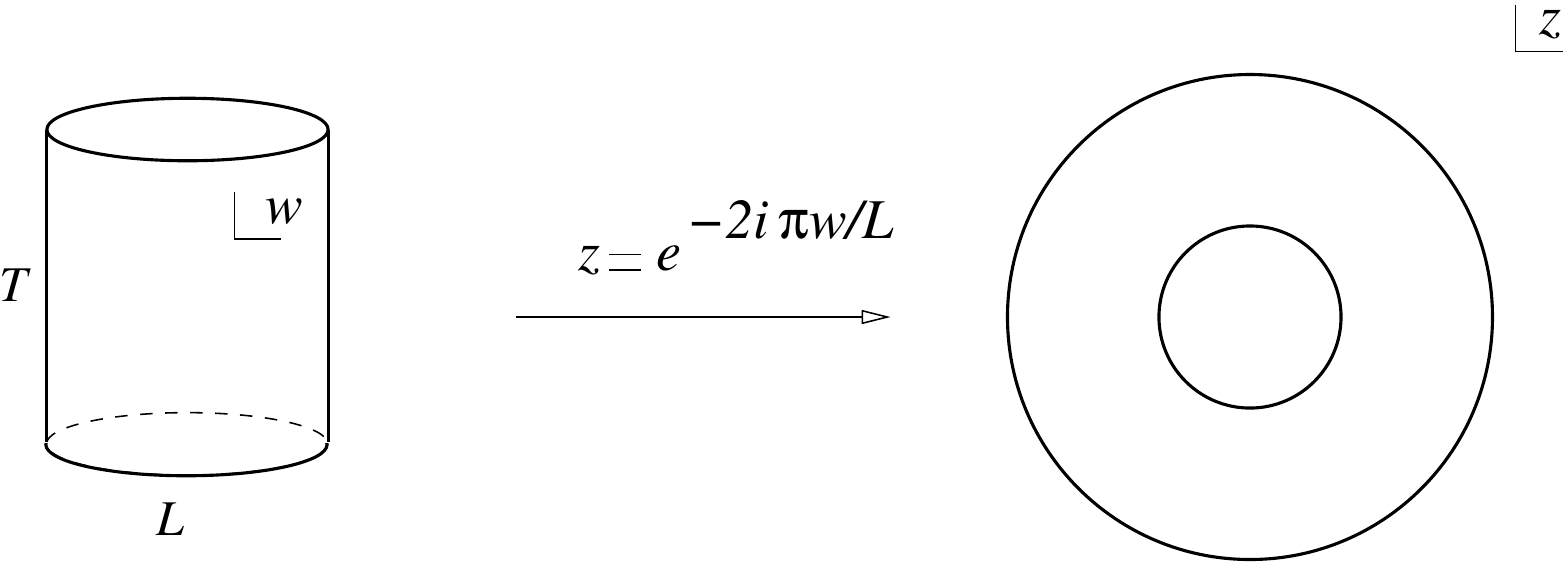}
\end{center}
\caption{Mapping from the cylinder to the annulus in the punctured plane: 
the two edges are identified for periodic boundary conditions, while they are
left independent in boundary CFT described in Section 4.}
\label{fig-map}
\end{figure}

\begin{figure}
\begin{center}
\includegraphics[width=11.5cm]{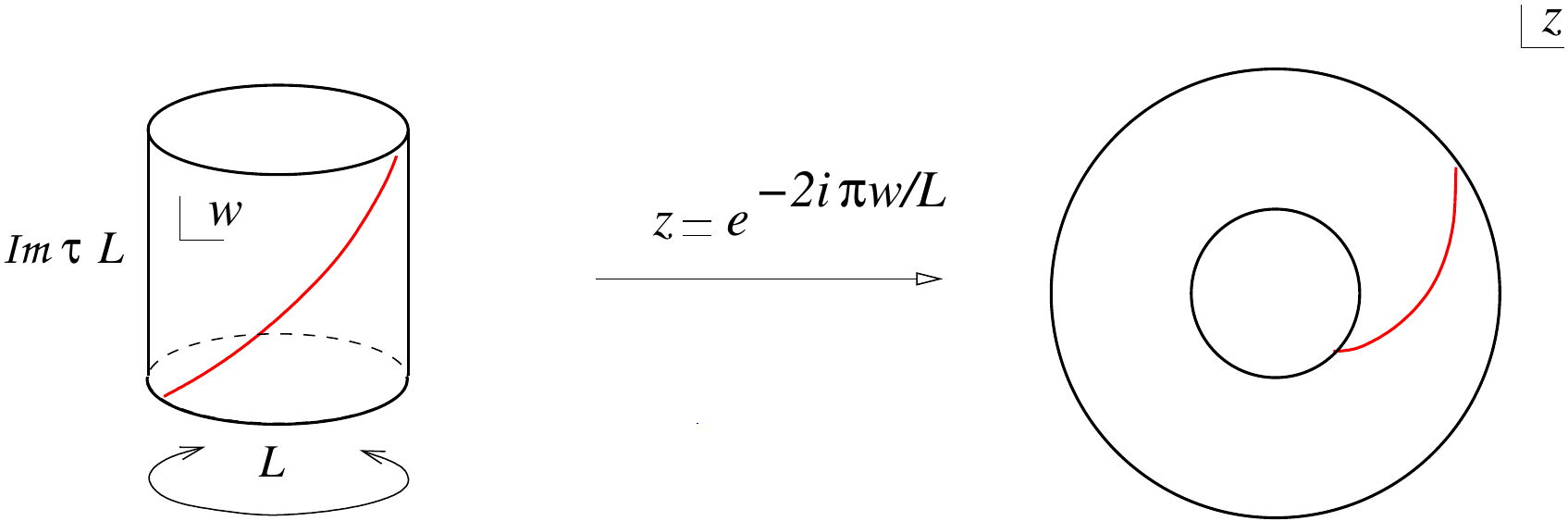}
\end{center}
\caption{Time evolution on the cylinder along an helix, and its image in 
the annulus.}
\label{fig2b}
\end{figure}

We start by considering the theory defined on a cylinder (see
Fig.\ref{fig-map}) of perimeter $L$ with a coordinate $w$; points
  of coordinates $w$ and $w+L$ are identified. This cylinder is
equivalent to the plane punctured at the origin, equipped with a
complex coordinate $z$, by means of the conformal mapping
$z=\exp(-2\pi i w/L)$.  On the cylinder, it is natural to think of the
Hamiltonian as the operator of translation along its axis (the
imaginary axis in $w$), or more generally along any helix, defined by its 
complex period
$\tau L$ in the $w$ plane, with $\Im m\,\tau>0$: ${\tau = i T/L}$ in the
particular case of Fig.\ref{fig-tm}, while the general case is
depicted in Fig. \ref{fig2b}.
The Virasoro generators of translations in $w$ and $\bar w$
(regarded as independent variables) are identified with
$L_{-1}^{{\rm cyl}}$ and  $\bar L_{-1}^{{\rm cyl}}$, therefore 
$H^{{\rm cyl}}= \tau L_{-1}^{{\rm cyl}}
+\bar\tau \bar L_{-1}^{{\rm cyl}}$, 
with $\bar\tau$ the complex conjugate of $\tau$.


Through the conformal mapping $w\mapsto z$, 
it is clear that these translations correspond
to dilatations and rotations in the plane.  Indeed, using the
transformation law of the energy-momentum tensor, one finds \cite{BCFT}: 
\be
\label{cyltopl}{ L_{-1}^{{\rm cyl}} = 
-{2\pi i\over L} ( L_0 -{c\over 24})}\ ,
\ee
where $L_0$ belongs to the Virasoro algebra of the plane and
the term $c/24$ comes from the conformal anomaly. 
In the continuum formulation, the transfer matrix is the exponential of 
the Hamiltonian, ${\bf T}= \exp(-\Delta t\ H)$ and therefore, the evolution
operator on a cylinder of length $(\Im m\,\tau) L$ is given by:
\be
\label{evol}
\exp\left(-H^{\rm cyl} L \right)=\exp\left[
2\pi i\tau \left(L_0-{c\over 24}\right) -
2\pi i \bar\tau\left(\bar L_0 -{c\over 24}\right)\right]\ .
\ee

We now introduce the partition function of the theory on a torus $\mathbb{T}$:
this is expressed by the trace of the evolution operator,
owing to the identification of the two ends of the cylinder,
\be
\label{Ztorus}
Z\left(\tau \right)=\tr_{\CH} \ e^{2\pi i [\tau (L_0-{c\over 24}) -\bar\tau
(\bar L_0-{c\over 24})] } \ .
\ee

Each irreducible representation $\CV_j$ of Vir (or of $\CA$) is graded
for the action of the Virasoro generator $L_0$: the spectrum of $L_0$
in $\CV_j$ is of the form $\{h_j,h_j+1,h_j+2, \cdots\}$, with
non-trivial multiplicities $m_n= \dim$(subspace of eigenvalue
$h_j+n$). It is thus natural to introduce a generating function of
these multiplicities, i.e. a function of a dummy variable $q$,
called the character of the representation $\CV_j$: 
\be
\label{chara}
\chi_{j}(q)=\tr_{\CV_{j}}\  q^{L_0-{{c\over 24}}}=q^{h_j-{{c\over 24}}}
\sum_{n=0}^\infty m_n q^n \ . 
\ee
These functions have been computed for several classes of chiral algebras
\cite{DFMS97}.

Using (\ref{hilbert}) and the definition of characters,
the trace in (\ref{Ztorus}) may be written as:
\be
\label{toruspf}
Z\left(\tau \right)=\sum_{j\bj}\ N_{j\bj} \ \chi_j(q)\ \chi_\bj (\bar q)\ ,
\qquad\qquad
q=e^{2\pi i\tau}\ ,\quad \bar q=e^{-2\pi i\bar\tau} \ . 
\ee
Let's stress that in these expressions, 
$\bar q$ is the complex conjugate of $q$, and therefore, $Z$
is a sesquilinear form in the characters.
Equation (\ref{toruspf}) shows that the torus partition function provides
a convenient way of encoding the structure of the RCFT Hilbert space 
(\ref{hilbert}).

\begin{figure*}
\begin{center}
\includegraphics[width=5cm]{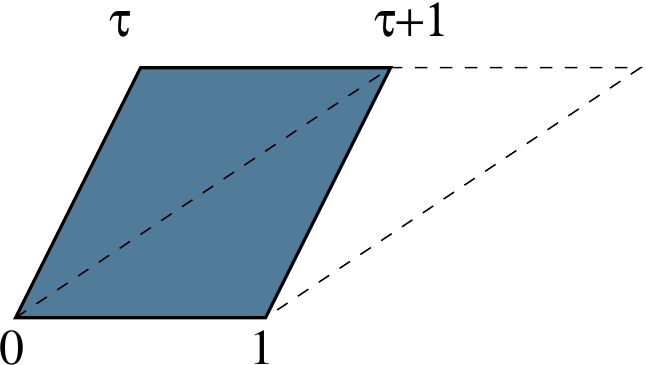}
\end{center}
\caption{Elementary cell in the $w$-plane describing the torus. The
$T$ modular transformation leads to the equivalent cell with 
parameter $\tau'=\tau +1$.}
\label{fig-tor}
\end{figure*}

The geometry of the torus $\mathbb{T}$
has been specified in (\ref{toruspf}) through
the modular parameter $\tau$, $\Im m\,\tau>0$; the two periods 
are $1$ and $\tau$, up to a global, irrelevant dilatation by $L$.
Equivalently, the torus may be regarded as the quotient of the complex
$w$-plane by the lattice generated by the two numbers $1$ and $\tau$:
\be
\label{torus}
\mathbb{T}= \IC/(\IZ \oplus \tau\IZ)\ , 
\ee 
in the sense that points are identified according to 
$w\sim w'=w+n+m \tau$, $n,m\in \IZ$.
This shows that there is a redundancy in the description of the 
torus: the modular parameters $\tau$ and $(c  + d\tau)/(a+b\tau)$ describe 
the same torus, for any transformation of the modular group: 
$M=\left({a\ b \atop c\ d}\right) \in PSL(2,\IZ)$, with
$a,b,c,d\in \mathbb{Z}$ defined up to a global sign and satisfying 
$ad-bc=1$. The group is generated by the two transformations:
$T: \tau \to \tau+1$ and $S: \tau \to -1/\tau$
\cite{Knopp70}: the first one is depicted in
Fig.\ref{fig-tor} and the second exchanges the two periods of the torus
generalizing the case in Fig.\ref{fig-tm}.

The partition function must be uniquely defined on the torus, 
independently of coordinate choices, and thus be invariant under 
modular transformations:
\be
Z \left( -\frac{1}{\tau} \right)= Z \left( \tau+1 \right) =
Z\left(\tau \right)\ .
\label{z-inv}
\ee 
These conditions, together with the expression (\ref{toruspf}) of $Z$ as 
a sesquilinear form in the characters, were introduced by 
Cardy \cite{Cardy86}. As we shall show in the following, 
they open a route to the classification of RCFTs.

In RCFTs, the finite set of characters $\chi_j$, $j\in \CI$,
transforms under modular transformations according to a unitary 
linear representation.
It suffices to give the action of the two generators of the modular group
acting on $\tau$, hence on $q=\exp 2\pi i\tau$,
\begin{eqnarray}
\tau \mapsto \tau'=\tau+1\ ,\qquad 
&& \chi_j(q')= e^{2\pi i (h_j-\frac{c}{24})} \chi_j(q)
 =\sum_{j'\in \CI} T_{jj'}\ \chi_{j'}(q)\ , 
\label{Tmod}\\
\tau \mapsto \tau'=-\frac{1}{\tau}\ , \qquad\quad
 && \chi_j(q')=\sum_{j'\in \CI} S_{jj'}\ \chi_{j'}(q)\ ;
\label{Smod}
\end{eqnarray}
$T$ and $S$ are unitary matrices \cite{Ver}.
It follows that RCFT partition functions (\ref{toruspf}) are modular
invariant if the (non-negative) integer multiplicity matrices 
$N_{j\bj}$ obey the conditions:
\be
SNS^\dagger =TNT^\dagger=N \ .
\label{n-cond}
\ee
When explicit computations of the partition function (path integral) 
are possible, the results should automatically be modular invariant.
In case of the strongly-interacting RCFTs, such derivations are not usually 
available: one can consider the algebraic construction 
described here, starting from the representations  
of the chiral algebra, and impose the modular invariance condition.
Let us stress that in this approach, modular invariance of $Z$ is
just one necessary condition for the consistency of the theory: 
there are additional requirements 
on other quantities, such as the crossing symmetry of the four-point
function on the plane \cite{BPZ84} \cite{MS89} (See \cite{Ver}).


\section{A-D-E Classification}

We thus are led to the following:\\
\noindent{\bf Classification Problem}. {
\sl For a given chiral algebra, find all possible Hilbert spaces 
(\ref{hilbert}), i.e. all possible sesquilinear forms 
(\ref{toruspf})
with non negative integer coefficients that are modular invariant, and such
that $N_{11}=1$.} 

The label $1$ refers to the identity representation (with $h_1=0$) and
the condition $N_{11}=1$ expresses the unicity of the ground state.  
As mentioned above, the finite set of characters of any RCFT, labelled by
$\CI$, supports a unitary representation of the modular group. This
implies that the diagonal combination of characters,
$Z=\sum_{i\in\CI} \chi_i(q) \ov{\chi_i(q)}$,
is always modular invariant for any RCFT.

Finding the general solutions for $Z$ in any RCFT is a well posed,
although involved, algebraic problem.  Complete results are known in
a few cases, in particular in several families of RCFT's based on the
$sl(2)$ algebra, for which a special feature, the ADE classification,
emerges.

\begin{center}
\begin{table*}
$$
\begin{array}{|c|c|c|}
\hline
{\rm level}& Z& {\rm diagram}
\\
\hline \hline
k\ge 0  
& \displaystyle{
\sum_{\lambda=1} ^{k+1} 
\left\vert\chi_{\lambda}\right\vert^2
}
&  A_{k+1} 
\\
\hline
k=4\rho\ge 4  
&  \displaystyle{
\sum ^{2\rho-1}_{\lambda \ {\rm odd}\ = 1}
\left\vert\chi_{\lambda}+\chi_{4\rho+2-\lambda}\right\vert^2
+2\left\vert\chi_{2\rho +1}\right\vert^2 
}
&  D_{2\rho+2} 
\\
\hline
k=4\rho -2\ge 6
& \displaystyle{
\sum ^{4\rho-1}_{\lambda \ {\rm odd}\ = 1}
\left\vert\chi_{\lambda}\right\vert^2 +
\left\vert \chi_{2\rho}\right\vert^2 +
\sum_{\lambda\ {\rm even} =2}^{2\rho-2} 
\left( \chi_{\lambda}\ov\chi_{4\rho-\lambda} +{\rm c.c.}\right)
} 
&  D_{2\rho+1}
\\
 \hline
k=10  
& \displaystyle{ 
\left\vert \chi_1  + \chi_7 \right\vert^2+
\left\vert \chi_4  + \chi_8 \right\vert^2 + 
\left\vert \chi_5  + \chi_{11} \right\vert^2 
}
& E_6 \\
\hline
k=16 &  
\begin{array}{c}
\displaystyle{
\left\vert \chi_1  + \chi_{17} \right\vert^2 + 
\left\vert \chi_5 + \chi_{13}  \right\vert^2 + 
\left\vert \chi_7  + \chi_{11} \right\vert^2  +
\left\vert \chi_9 \right\vert^2  }
\\ 
\displaystyle{
\qquad +\left[( \chi_3+\chi_{15} ) \ov\chi_9  +{\rm c.c.}  \right] }
\end{array} 
&  E_7 \\
\hline
k=28 &  \displaystyle{
\left\vert \chi_1  +\chi_{11} + \chi_{19} + \chi_{29} \right\vert^2
+\left\vert \chi_7  +\chi_{13} + \chi_{17} + \chi_{23} \right\vert^2  
}
&  E_8  \\
\hline
\end{array}
$$
\caption{List of modular invariant partition functions of
$\slh(2)$ RCFTs: $\chi_\lambda$ are characters
of representations of the affine algebra at level $k$.
The last column shows the associated ADE Dynkin diagram.}
\label{su2-table}
\end{table*}
\end{center}


\subsection{The $\slh(2)$ cases.}

For ${\cal A}=\slh(2)$, the $sl(2)$ affine Kac-Moody  algebra
at a non negative integer value of the ``level'' $k$ (its central charge), 
the set of possible ``integrable'' representations is labelled by an integer or
half-integer $j$ (the spin of the representation of the ``horizontal
algebra''), subject to $ 0\le j \le k/2$, or more conveniently
by the integer $\lambda= 2j+1$, $1\le \lambda\le k+1$ \cite{KM}. 
The matrix of modular transformations in (\ref{Smod}) reads:
\be
S_{\lambda\lambda'}=\left(\frac{2}{k+2}\right)^{1/2}
\sin\left(\frac{\pi\lambda\lambda'}{k+2}\right)\ ,
\ee
the Virasoro central charge and the conformal weights are,
\be
c=\frac{3k}{k+2}\ ,\qquad\quad
h_\lambda=\frac{\lambda^2-1}{4(k+2)} \ .
\ee

\begin{table*}
\begin{center}
\begin{tabular}{|lcll|}
\hline
$G$ &  diagram &\qquad $h$  & \qquad  exponents $\ell_n$
\\
\hline\hline
$A_n$& \includegraphics[trim=0 0 0 -20,width=4cm]{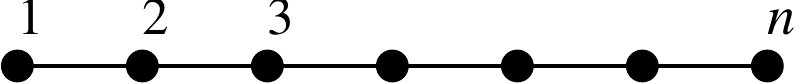}
& \qquad   $n+1$  &\qquad $1,2,\cdots,n$ 
\\
$D_{n+2}$ &\includegraphics[trim=0 45 0 -10,width=4cm]{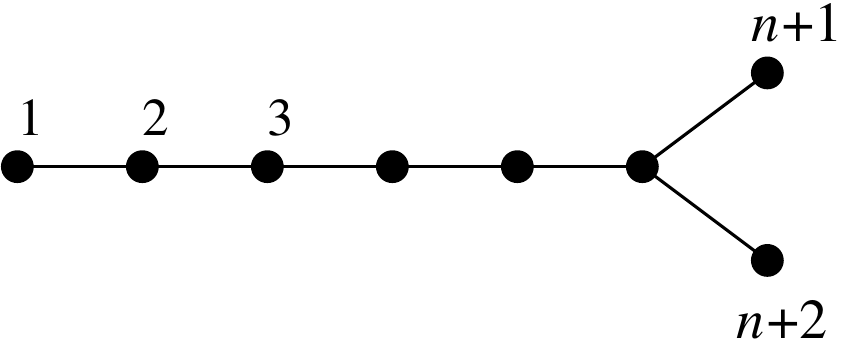}
&\qquad  $2(n+1)$ &\qquad $1,3,\cdots,2n+1,n+1$ 
\\
$E_6$ & \includegraphics[trim=0 10 0 -10,width=2.5cm]{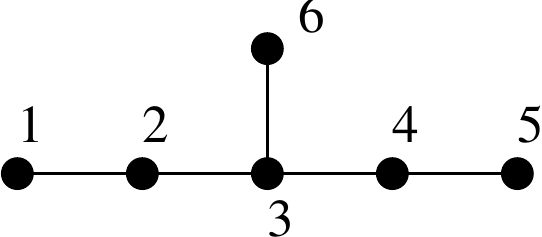} &\qquad 12  
&\qquad  $1,4,5,7,8,11 $
\\
$E_7$ & \includegraphics[trim=0 10 0 -10,width=3cm]{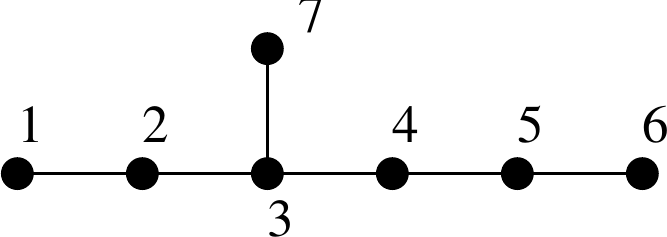} 
&\qquad 18& \qquad $1,5,7,9,11,13,17$  
\\
$E_8$ &\includegraphics[trim=0 10 0 -10,width=3.5cm]{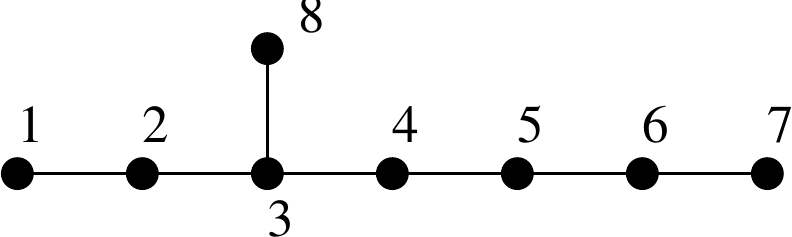}
&\qquad  30   & \qquad $1,7,11,13,17,19,23,29$ 
\\ 
\hline
\end{tabular}
\caption{ADE Dynkin diagrams with Coxeter numbers $h$ and exponents
$\ell_n$.}
\label{ade-table}
\end{center}
\end{table*}

The study of modular invariance of partition functions leads to the
determination of all possible matrices $N_{\lambda,\bar\lambda}$ 
in (\ref{toruspf}); they are displayed in Table \ref{su2-table}.
The result was first conjectured in \cite{CIZ87a} and then proved in
\cite{CIZ87b}, \cite{K87}; partial results had been previously obtained by
several authors, including \cite{Cardy86} and \cite{Gep87}.
The analysis was made in two steps:
i) the derivation of the complete list of sesquilinear invariants
forms irrespectively of the positivity of their coefficients;
ii) the imposition of positivity. 
The proof is rather long and cannot be described here: let us just give a
clue.  Given that the $S$ matrix is a discrete Fourier transform, one is
dealing with the algebra of numbers $\lambda$ defined modulo $n=2(k+2)$: one
finds that there is one independent modular invariant for each number $\omega$
solution of $\omega^2=1$ mod $2n$ \cite{CIZ87a} \cite{GQ87}. 
It is given by a sort of permutation matrix, 
$N_{\lambda,\lambda'}=\delta_{\lambda,\omega\lambda'\ {\rm mod}\ n}$,
that obeys the conditions (\ref{n-cond}).
The range of indices of the characters, however, is only $n/2$ rather
than $n$, with the following periodicity property:
$\chi_\lambda=-\chi_{\lambda+n/2}=\chi_{\lambda+n}$;
therefore, most of these solutions actually correspond to sesquilinear forms
(\ref{toruspf}) with some negative coefficients.
A deeper analysis then obtained the subset of positive definite 
ones that are shown in Table \ref{su2-table} \cite{CIZ87b}.
More recently, a simpler method was found that directly searches for
the positive invariants \cite{Gan00}.
 
The list of modular invariants in Table \ref{su2-table} exhibits a remarkable
feature: a one-to-one correspondence with the ADE Dynkin diagrams, which are
best known for the classification of simply-laced simple Lie algebras (See
Table \ref{ade-table}) \cite{Hum72}.  The relation is as follows: if we
 analyze the diagonal terms of expressions in Table \ref{su2-table},
their labels $\lambda$ turn out to be the 
``Coxeter exponents'' $\ell_n$ of the Dynkin diagrams; 
these are  given in Table \ref{ade-table} and will be defined shortly.
For example, in the $E_7$ modular invariant of Table 1, the diagonal terms are
labelled by odd integers from 1 to 17, excluding 3 and 15, which matches the
$E_7$ exponents of Table 2.
Moreover, the Coxeter number $h$, another characteristic of the diagrams, 
is related to the $\slh(2)$  level by $h=k+2$.

We recall that the Dynkin diagrams encode the geometry of root
systems $\{\alpha_a\}$ of simple Lie algebras \cite{Hum72}: the
simply-laced ones have roots of equal length and the Cartan matrix,
$C_{ab}=(\alpha_a,\alpha_b)$, is expressed in terms of the adjacency
matrix $G$ of the diagram, $C_{ab}=2 I -G_{ab}$ (the matrix $G$ is
defined for any graph as $G_{ab}=1$ if sites $(a,b)$ are connected and
zero otherwise).
For these diagrams, the eigenvalues of the adjacency
matrix $G$ are of the form $2\cos (\pi \ell_n/ h)$, where 
$h$ is the Coxeter number, and $\ell_n$  the Coxeter exponents.
These numbers take rank($G$) values (the number of vertices of $G$),
ranging  between $1$ and $h-1$, with possible multiplicities.
The exponents also express other properties: for example, when shifted
by $1$, they give the degrees of all invariant polynomials of the Lie
algebra: the quadratic Casimir invariant ($\ell =1$) and the higher
invariants ($\ell>1$).  The Coxeter number and exponents are listed in
Table \ref{ade-table}.


\subsection{Minimal conformal models}

The ``minimal models'' introduced in Ref \cite{BPZ84} are the only
RCFT with values  $c<1$ of the central charge of the Virasoro algebra:
\be
c\left(p,p'\right)=1-\frac{6(p-p')^2}{pp'}\ , 
\qquad p,p'\in \IZ^+, \quad (p,p')=1,
\label{mm-c}
\ee
($p$ and $p'$ are positive coprime integers).
The highest weights of representations take values in 
the ``Kac table'' \cite{CFT}: 
\be 
h_{rs}=\frac{(rp-sp')^2-(p-p')^2 }{4 p p'},
\qquad 1\le r \le p'-1\ ,\quad 1\le s \le p-1\ ,
\label{kac-val}
\ee
quotiented by the equivalence $(r,s)\sim (p'-r,p-s)$. 
In Ref \cite{CIZ87a}, the partition functions of minimal models 
were shown to also follow the ADE scheme: the results are displayed 
in Table \ref{mm-table}. 
In this case, the partition functions are associated to pairs of diagrams:
one of them is always of the $A$ type, because one of the Coxeter numbers
$(p,p')$ is odd.
Besides the diagonal invariants, there are two infinite series,
$(D,A_{p-1})$ and $(A_{p'-1},D)$, and six exceptional cases,
$(E,A)$, $(A,E)$.
The similarity between the two ADE classifications can be traced back to
the relation between Virasoro representations and 
affine $\slh(2)$ algebra representations \cite{KM}, given by the coset
construction: ${\rm Vir}\sim \slh(2)_{p'}\otimes \slh(2)_{p-p'}/\slh(2)_{p}$
(for $p=p'+1$) \cite{GKO86}\cite{DFMS97}.

\begin{table*}
\begin{center}
$$ \begin{array}{|c|c|c|}
\hline
p' & Z &{\rm diagrams}
\\
\hline\hline
&  
\displaystyle{\frac{1}{2} \sum_{r=1} ^{p'-1} \sum_{s=1} ^{p-1}  
\left\vert\chi_{rs}\right\vert^2 }
&  
(A_{p'-1},A_{p-1})
\\
\hline
\displaystyle{p'=4\rho+2\atop \rho \ge 1}   
&  
\displaystyle{\frac{1}{2}  \sum_{s=1} ^{p-1}\left\{ 
\sum ^{2\rho-1}_{r\ {\rm odd}\ = 1}
\left\vert\chi_{r,s}+\chi_{4\rho+2-r,s}\right\vert^2+
2\left\vert\chi_{2\rho +1,s}\right\vert^2\right\} } 
&  
(D_{2\rho+2},A_{p-1}) 
\\
\hline
 \displaystyle{{p'=4\rho \atop \rho\ge 2}} 
 & 
\displaystyle{
\frac{1}{2} \sum_{s=1} ^{p-1}\left\{ \sum ^{4\rho-1}_{r \  {\rm odd}
\ = 1}\vert\chi_{r,s}\vert^2
 +\vert \chi_{2\rho,s}\vert^2 +\sum_{r\ {\rm even} =2}^{2\rho-2} 
\left( \chi_{r,s}\ov\chi_{4\rho-r,s} +{\rm c.c.}\right)\right\}}  
&  
(D_{2\rho+1},A_{p-1})
\\
 \hline
 p'=12  
& \displaystyle{\frac{1}{2} \sum_{s=1} ^{p-1}\left\{ 
\left\vert \chi_{1,s}  + \chi_{7,s} \right\vert^2+
\left\vert \chi_{4,s}  + \chi_{8,s} \right\vert^2 + 
\left\vert \chi_{5,s}  + \chi_{11,s} \right\vert^2 \right\}} 
&  
(E_6,A_{p-1}) 
\\
\hline
p'=18 
&  
\displaystyle{\frac{1}{2}  \sum_{s=1} ^{p-1} \left\{
\left\vert \chi_{1,s} + \chi_{17,s} \right\vert^2 + 
\left\vert \chi_{5,s} + \chi_{13,s}  \right\vert^2 +
\left\vert \chi_{7,s}  + \chi_{11,s} \right\vert^2  +
\left\vert \chi_{9,s} \right\vert^2 \right.}
\atop \displaystyle{\left.\qquad +
\left[ \left( \chi_{3,s}+\chi_{15,s} \right) \ov\chi_{9,s} +{\rm c.c.}  
\right]\right\}}  
&  
(E_7,A_{p-1}) 
\\
\hline
 p'=30  
& 
\displaystyle{\frac{1}{2}  \sum_{s=1} ^{p-1} \left\{ 
\left\vert \chi_{1,s}  +\chi_{11,s} + \chi_{19,s} + \chi_{29.s} \right\vert^2
+\left\vert \chi_{7,s } +\chi_{13,s} + \chi_{17,s} + \chi_{23,s} \right\vert^2
\right\} } 
&  
(E_8,A_{p-1})  
\\
\hline
\end{array}$$
\caption{List of modular invariant partition functions of minimal
models with $c(p,p') <1$: $\chi_{r,s}$ are characters of Virasoro
representations with highest weight $h_{rs}$ in (\ref{kac-val}).
Each invariant, but the first,  also occurs for $p\leftrightarrow p'$.}
\label{mm-table}
\end{center}
\end{table*}

The results in Table \ref{mm-table} amount to a classification of 
the operator contents of all rational conformal theories with $c<1$;
we see that more than one consistent set of primary fields is possible 
for the same central charge (\ref{mm-c}), leading to 
quite different theories.
These correspond to independent universality classes of critical phenomena
in statistical mechanics, because the (relevant) primary fields characterize 
the manifold of perturbations around the fixed point.
Let us mention some examples among the unitary minimal models
corresponding to $(p,p')=(m+1,m)$ or $(m,m+1)$, with $m=3,4,\dots$.
In the $(A_{m-1},A_{m})$ series, the $(A_2,A_3)$ model is the Ising
model, $(A_3,A_4)$ is the tricritical Ising and the higher models are
their restricted solid-on-solid (RSOS) generalizations
\cite{ABF84} \cite{Huse84} \cite{Pas87}.
The simplest non-diagonal partition functions are: $(A_4,D_4)$ corresponding
to the 3-state Potts model and $(D_4,A_6)$ its tricritical version
\cite{Cardy86}.

Related ADE classifications of partition functions were found for the
minimal superconformal theories \cite{Cap87}, the $\IZ_k$
parafermionic theories \cite{GQ87} and the $\CN=2$ superconformal theories
\cite{Gan97}, all being connected with the
$\slh(2)$ affine algebra by coset constructions.

 The classification of partition functions has also been completed for
the case of the affine $\slh(3)$ models \cite{Gan94}; the
interpretation of the results in terms of generalized graphs has been
considered in \cite{Kos88} \cite{DFZ90}. Partial results have been obtained for
other, higher rank algebras (for a review see \cite{Gan99}).


\section{Partition functions on the annulus with 
boundary conditions}

 The ADE classification of modular invariants described in the previous
section does not directly use, or relates to, properties of Lie algebras 
and the associated geometry of root lattices.
However, a closer connection has been found in the corresponding
classification of partition functions with a boundary \cite{BPPZ00}.

As discussed in \cite{BCFT}, the partition function
on the annulus (cf. Fig~\ref{fig-map}), with given 
conformal-preserving boundary conditions, is expressed in
terms of the conformal characters (\ref{chara}) of a single chiral algebra:
\be
Z_{a|b}(q)=\sum_{i\in\cal I}\ n_{ia}^{\ b}\ \chi_i(q)\ ,
\label{Zab}
\ee
where $q=\exp\left(-\pi L/T \right)$ in the notation of Fig.\ref{fig-map}.
In this equation, the $a,b$ indices label the boundary conditions on the
two edges of the annulus: as in the torus expansion (\ref{toruspf}), 
the integers $n_{ia}^{\ b}$ count the multiplicities of chiral
representations pertaining to the Hilbert space of states compatible
with the boundary conditions.
In RCFTs, the number of boundary conditions is also finite,
and a lot of information about the theory in the bulk may be obtained
from the study of boundary states.

We also need to introduce the ``fusion rules'' and ``fusion algebra''
\cite{Ver} \cite{DFMS97}: they give the selection rules for the
product of representations of chiral algebras of RCFTs, inherited from
the operator product expansion of quantum field theory.  It is natural
to decompose the fusion of two representations labelled by $i,j\in
\CI$ into other representations, as follows:
\be
{\cal V}_i\bullet {\cal V}_j = \bigoplus_{k\in\CI}\ 
\CN _{ij}{}^k\ {\cal V}_k\ ,
\label{fusion}
\ee 
thus defining the multiplicities, or ``fusion coefficients'', $
\CN _{ij}{}^k$. Due to the properties of the operator product,
the $(\CN_i)_j^{\ k}$ matrices yield an associative and commutative
algebra over the integer numbers called fusion algebra.

There is a remarkable formula, due to Verlinde \cite{Ver88}\cite{Ver},
expressing these multiplicities in terms of the unitary modular matrix
$S$ defined in (\ref{Smod}): 
\be \CN_{ij}{}^k=\sum_{\ell\in\CI}
\frac{S_{i\ell}\ S_{j\ell}\ \ov{S}_{k\ell}}{S_{1\ell}} \ .
\label{Verl}
\ee 
This general result in RCFT follows from consistency conditions
relating multipoint amplitudes on Riemann surfaces \cite{MS89}.

In the annulus geometry, the partition function can also be expressed
in two alternative (and different) ways corresponding to evolutions in
the two orthogonal directions (cf. Fig. \ref{fig-tm}), one of whose is
(\ref{Zab}).  Their comparison yields Cardy's equation of modular
covariance on the annulus \cite{BCFT} \cite{Cardy89}.
Supplemented by some technical assumptions of orthogonality and
completeness of boundary conditions \cite{PSS95}, 
this equation implies that the
matrices of multiplicities, $n_i=\left(n_{i}\right)_a^{\ b}$, in
(\ref{Zab}) must form a representation of the fusion algebra,
\be
{n_i\;n_j=\sum_{k\in\CI} \CN_{ij}{}^k\,n_k \ }\ .
\label{fus-alg}
\ee
The study of (orthogonal complete) boundary conditions is thus given 
by the study of non-negative integer valued matrix representations 
(or ``nimreps'') of the fusion algebra \cite{BPPZ00}.
In the case of diagonal torus invariants, one actually has
$(n_i)_j^{\ k}=(\CN_j)_j^{\ k}$, and there is a complete correspondence 
between the bulk and boundary sectors of the theory.
For general, non-diagonal torus partition functions, one finds that
the eigenvalues of $n_i$ are the same as those of $\CN_i$, and
are of the form $S_{ij}/S_{1j}$, owing to (\ref{Verl}); 
however, the $j$ label can only take the values corresponding to 
the diagonal terms, for which $N_{jj}\neq 0$
in the expression of the torus partition function (\ref{toruspf}). 

In the case of $\slh(2)$ theories, the study of boundary conditions
can be carried out completely.  
The fusion algebra is generated by the
first non trivial matrix $\CN_2$ (corresponding to spin $1/2 $).
From section 3.1, its eigenvalues are of the form 
$ S_{2\ell}/S_{1\ell}=\sin(2\pi \ell/h) 
/ \sin  (\pi \ell/h)= 2\cos (\pi \ell/h)$, 
thus in the interval $]-2,2[$. 
This property is shared by the matrix $n_2$ of multiplicities in
(\ref{Zab}), which also generates the whole set $\{ n_i\}$. 
The study of boundary conditions thus reduces to finding all
matrices $n_2$ with that spectral property.

A theorem (see \cite{GHJ89}) states that
the symmetric matrices with non negative integer entries
and eigenvalues between $-2$ and $2$ are the adjacency matrices
$G_{ab}$ of the ADE graphs of Table \ref{ade-table}. 
Therefore, the ADE classification of $\slh(2)$ theories, obtained through the
imposition of their consistency on a torus, reappears through the
consistency of their possible boundary conditions.
In particular, the occurrence of Coxeter exponents as labelling diagonal terms
in the torus partition functions receives a natural explanation
 (see Table \ref{su2-table}).
We remark that the same characterization of ADE graphs also
occurs for the Cartan matrix in the classification of Lie algebras; 
this provides a more direct link between the two problems.

(For completeness, we observe that the set of $\slh(2)$
boundary conditions, i.e. of fusion algebra representations, is not
completely equivalent to that of torus partition functions.
Actually, the theorem above provides additional fusion matrices $n_2$, 
associated to ``tadpole'' graphs $T_n=A_{2n}/\IZ_2$ \cite{BPPZ00}, 
but they do not correspond to any modular invariant $Z$.
One can show by inspection that a multiplicity matrix 
$N_{\lambda\lambda'}$ in (\ref{toruspf}), 
whose diagonal terms would be labelled by tadpole exponents, 
could not satisfy the invariance conditions (\ref{n-cond}).
Still, the study of boundary conditions leads to a neat appearence of the
ADE scheme.)


\section{Physical realizations of A-D-E partition functions}

There are two classes of CFTs for which the ADE classification appears
from a more physical standpoint: the $c<1$ (unitary) minimal
models, and the minimal $\CN=2$ superconformal field theories ($c<3$)
(and their topological cousins).

In the former case, the $c<1$ minimal models admit a realization as
integrable statistical systems on a lattice \cite{Pas87}.  In these
solid-on-solid models, the patterns of heights
are given by paths on a graph: by demanding
that the configuration space supports a representation of the
Temperley-Lieb algebra (a quantum deformation of the symmetric group
algebra and a known way to achieve Yang-Baxter integrability
\cite{YB}) and that the model is at the critical point, one is led to
graphs whose adjacency matrix has eigenvalues between $-2$ and $2$,
hence of ADE type.
This construction of minimal models was found at the
same time of the classification of partition functions, thus providing
a useful physical implementation of it.  The lattice model based on a
Dynkin diagram $G$ of ADE type, with Coxeter number $h$, is mapped in
the continuum limit onto the minimal CFT with $(A_{h-2},G)$ modular
invariant, while the $(G,A_{h})$ modular invariant describes a
tricritical version of the same lattice model \cite{Kos92} \cite{Roche92}
\cite{WNS92}.

The minimal $\CN=2$ superCFTs admit a Lagrangian description, involving 
chiral superfields and a holomorphic superpotential
$W$ \cite{LVW89} \cite{Mar89}. 
Due to supersymmetry, the form of $W$ is protected by
nonrenormalization theorems; therefore, the semiclassical
Landau-Ginsburg theory \cite{LG} of critical points as stationary
points of the potential $W$ holds at the full quantum level.
For example, the simpler $m$-th critical point is described in terms
of a single chiral field $\Phi$ and the potential $W=\lambda\Phi^{m+1}$.
A perturbation that moves the theory off 
the critical point and eventually lowers the degree of criticality
is represented by:
\be
W\left(\Phi\right)=\lambda \Phi^{m+1} + \mu_k \Phi^k\ , \qquad k=1,\dots,m-1 .
\label{LG-pot}
\ee
Indeed, the coupling $\mu_k\in[0,\infty]$ is relevant and describes 
the renormalization-group flow from the $m$-th critical point to 
the $(k-1)$-th one.
The phase diagram around multicritical points is described by these
perturbations that match Wilson's analysis of relevant and marginal
scaling fields; the higher the degree of criticality, the larger the
number of these fields.

In this setting, one can make contact with another occurrence of the
ADE scheme, that of Arnold's singularities \cite{AGZV85}.  This is
the study of critical points of polynomials $W(x_1,x_2, \cdots,x_p)$,
i.e. $\partial W/\partial x_i|_0=0$ for all $i=1,\cdots,p$,
that are identified up to diffeomorphisms of the $x_i$ coordinates.  As in
(\ref{LG-pot}), each singularity has an associated vector space of
``genuine'' deformations, that cannot be removed by
reparametrizations: they may keep the same degree of the singularity
or lower it, namely keep the vector space unchanged or reduce it,
respectively.
Deformations of the former type are called ``moduli''
and the ``simple'' singularities have none of them.  A
theorem states that simple singularities for any number $p$ of
coordinates fall into the ADE scheme: they
are actually given by the ``Kleinian polynomials'' $W$
that are listed in Table \ref{k-table} 
(discarding the quadratic terms) (\cite{AGZV85}.

Arnold's study of singularities can be used to classify the stationary
points of Landau-Ginsburg potentials: the reparametrizations are
redefinitions of the effective field(s) that have no physical effect,
and the moduli are deformations preserving the critical degree, i.e.
marginal perturbations.  The simple singularities correspond to isolated
critical points, that is, without marginal operators.
Therefore, the ADE scheme classifies the isolated critical
points of $\CN =2$ Landau-Ginsburg superpotentials and matches the
series of superconformal minimal models \cite{LVW89} \cite{Mar89}. 
This description applies to the ``chiral sector'' of the theory
\cite{Gan97}.

Let us notice that the classifications of $\CN =2$ and $\CN=0$
  (Virasoro) minimal models are very similar: the former involve one
  ADE scheme (through the $\slh(2)$ classification, Table
  \ref{su2-table}), the second a double scheme (Table
  \ref{mm-table}). Indeed, several results have shown that a
qualitative Landau-Ginsburg description of Virasoro minimal models is
possible, in spite of renormalization changing the
  conformal dimensions of fields \cite{Zam86} \cite{LC87}
(\cite{CDZ04} discuss the case with boundaries). 
However, a detailed correspondence with ADE singularities
cannot be found as in the supersymmetric case. 
Still the Landau-Ginsburg description, supplemented by the
Arnold study of simple singularities, provides a qualitative physical 
picture for the ADE schemes found in the minimal CFTs.

We finally mention that rational CFTs find several applications in
low-dimensional condensed-matter systems, like spin chains and quantum
wires in $(1+1)$ dimensions \cite{Tsvelik} and quantum Hall effect in
$(2+1)$ dimensions \cite{Wen}.  In the latter case in particular,
modular invariant partition functions have been found in \cite{CGT01}
\cite{CVZ09} and the RCFT methods of section 4 have been exploited in
\cite{QC}.

\begin{table}
$$
\begin{array}{|c||ccccc|}
\hline
\Gamma & \CC_n &\CD_n & \CT& \CO & \CI
\\
|\Gamma| & n & 4n& 24 & 48 & 120
\\
W& X^{n} -Y Z & X^{n+1}+ XY^2 +Z^2 & X^4+Y^3+Z^2 & X^3+X Y^3 +Z^2 &
X^5+Y^3+Z^2 
\\
G & A_{n-1} & D_{n+2} & E_6 & E_7 & E_8
\\
\hline
\end{array}
$$
\caption{ Finite subgroups $\Gamma$ of $SU(2)$, their orders $|\Gamma|$, 
the Kleinian polynomials $W$ and the associated Dynkin diagram.
Arnold's simple singularities are the polynomials without quadratic terms.}
\label{k-table}
\end{table}


\section{Elements of ADE-ology}

The occurrence of the ADE scheme in the solution of modular invariant
partition functions is a rather remarkable result, but we saw it can be 
traced back to a simple fact: certain discrete equations admits a
simple, yet non-trivial set of solutions, which are in one-to-one
correspondence with Dynkin diagrams.
Besides the classification of Lie algebras and simple singularities, 
the ADE scheme also appear in other seemingly unrelated problems of 
physics and mathematics.
It is rather interesting to find analogies and correspondences among them.

Other occurrences of the ADE scheme are found in:

\begin{tabular}{p{.5cm}p{14cm}}
{(i)} & finite reflection groups of crystallographic and of
simply-laced type \cite{Hum90};\\
{(ii)} & finite subgroups of $SO(3)$ or of $SU(2)$, and the associated
Platonic solids;\\
{(iii)} & Kleinian singularities \cite{Slo83};\\
{(iv)} & finite type quivers  \cite{Gab72}; \\
{(v)} & algebraic solutions to the hypergeometric equation 
\cite{Hille76};  \\
{(vi)}& subfactors of finite index \cite{Jones83};\\ 
\end{tabular}

and the list is probably not exhaustive.

Let us briefly describe some of  these classifications
and point to the relevant literature.
 
We already introduced the ADE Dynkin diagrams that arise in the
classification of simple Lie algebras by Killing and Cartan \cite{Hum72}.  
In the case of crystallographic groups, the Dynkin diagrams describe the
geometry of the hyperplanes of reflection and of the simple root
vectors orthogonal to them.  The product of the reflections, 
$S_a(\beta)=\beta - 2 \frac{(\beta,\alpha_a)}{(\alpha_a,\alpha_a)}\alpha_a$, 
over all positive simple roots
$\alpha_a$ defines the Coxeter element, unique up to conjugation, whose
eigenvalues are $\exp( 2i\pi \ell_i/h)$, with $\ell_i$ running over
the exponents defined above in section 3.1 (Table \ref{ade-table}).

Finite subgroups of $SU(2)$ form two infinite series and three
exceptional cases: the cyclic groups $\CC_n$, the binary dihedral groups
$\CD_n$, the binary tetrahedral group $\CT$, the binary octahedral group $\CO$
and the binary icosahedral group $\CI$. 
It is natural to label them by ADE as in Table \ref{k-table}.
A related classification is that of the regular solids in three dimensions:
this may be the oldest ADE classification, since it goes back to the
school of Plato; here, $E_6$ is associated with the symmetry group
of the tetrahedron, $E_7$ with the group of the octahedron or of the
cube, and $E_8$ with  the group of the dodecahedron or of the icosahedron.
The cyclic and dihedral groups may be
thought of respectively  as the rotation invariance group of a pyramid 
and of a prismus of base a regular $n$-gon, but those are not
regular Platonic solids. The McKay correspondence provides a direct
relation between subgroups and Dynkin diagrams \cite{McKay80}
\cite{Zuber00}.

Kleinian singularities (Table \ref{k-table}):
 Let $\Gamma$ be a finite subgroup of $SU(2)$. It acts on
$(u,v)\in\IC^2$. The algebra of $\Gamma$-invariant polynomials in $u,v$
is generated by three polynomials $X,Y,Z$ subject to one relation
$W(X,Y,Z)=0$. The quotient variety $\IC^2/\Gamma$ is
parameterized by these polynomials $X,Y,Z$ and is thus 
embedded into the hypersurface $W(x,y,z)=0$, $(x,y,z)\in \IC^3$.  
This variety has an isolated singularity at the origin \cite{Slo83}.

In many cases, the classification follows from the spectral condition
for symmetric matrices discussed in section 4.
In some others, however, the key point is the determination
of triplets of integers $(p,q,r)$ such that:
\be
{1\over p}+{1\over q}+{1\over r}>1 \ .
\ee
These are:

\be
\begin{array}{|l||lllll|}
\hline
G & A_{2n-1} & D_{n+2}& E_6 & E_7 & E_8
\\
(p,q,r) &(1,n,n) &(2,2,n) &(2,3,3) & (2,3,4) &(2,3,5)
\\
\hline
\end{array}
\label{pqr}
\ee

Note that the ADE list above includes all the solutions except $(p=1, q\ne r)$.
Note  also that for the $D$ and $E$ cases, these integers give the length 
(plus one) of the three branches of the Dynkin diagram counted from the vertex 
of valency three.

This short account does not exhaust all the facets of this fascinating
subject. The ADE scheme also manifests itself in related problems,
namely:
\begin{itemize}
\item 
the construction by Ocneanu of topological invariants \`a la Turaev-Reshetikhin
 with the methods of operator algebras, their ADE classification, 
 and their connection with Jones classification of subfactors \cite{Ocn};
\item 
the discussion of the minimal topological (``cohomological'')
  field theories, that may be regarded as twisted versions of the
  $\CN=2$ superconformal theories mentioned above: in these theories,
  the (genus 0) three-point functions satisfy the so-called
  Witten-Dijkgraaf-Verlinde-Verlinde equations \cite{Wit90} \cite{DVV91}, 
  for which Dubrovin has shown the appearance of monodromy groups
  generated by reflections \cite{Dub}.
\end{itemize}

Finally, the ADE scheme not only controls the spectrum of the minimal
conformal theories, but also contains some important information about
their Operator Product Algebra (OPA).  It can be proved that the
structure constants of the OPA of all the minimal theories may be
determined from those of the diagonal theories in terms of the
eigenvectors of the ADE Dynkin diagrams \cite{Pas87} \cite{PZ95}.

\bigskip

{\bf Acknowledgments}

Our first thought goes to a great friend, Claude Itzykson.  We
would like to thank all colleagues that shared their insight and
enthusiasm on this subject with us; in particular, Michel Bauer,
Roger Behrend, Denis Bernard, John Cardy, Philippe Di Francesco, Terry Gannon, 
Lachezar Georgiev, Doron Gepner, Victor Kac,
Ivan Kostov, Vincent Pasquier, Paul Pearce, Valentina Petkova,
Hubert Saleur, Ivan Todorov and Guillermo Zemba.


\end{document}